\documentclass[a4paper,fleqn,usenatbib,letters]{mnras}

\usepackage[T1]{fontenc}
\usepackage{ae,aecompl}
\usepackage{pdflscape}

\usepackage{graphicx}	
\usepackage{amsmath}	

\usepackage{amssymb}	

\title[3D Structure of Upper Scorpius]{Three-dimensional structure of the Upper Scorpius association with the Gaia first data release}

\author[P. A. B. Galli, I. Joncour \& E. Moraux]{
Phillip A. B. Galli,$^{1}$\thanks{E-mail: phillip.galli@iag.usp.br}
Isabelle Joncour$^{2,3}$ and
Estelle Moraux$^{2}$
\\
$^{1}$Instituto de Astronomia, Geof\'isica e Ci\^encias Atmosf\'ericas, Universidade de S\~ao Paulo, Rua do Mat\~ao, 1226,\\ Cidade Universit\'aria, 05508-900, S\~ao Paulo - SP, Brazil\\
$^{2}$Univ. Grenoble Alpes, CNRS, IPAG, 38000, Grenoble, France\\
$^{3}$Department of Astronomy, University of Maryland, College Park, MD 20742, USA
}

\date{Accepted 2018 March 02. Received 2018 March 02; in original form 2017 November 21}

\pubyear{2018}

\begin{document}
\label{firstpage}
\pagerange{\pageref{firstpage}--\pageref{lastpage}}
\maketitle

\begin{abstract}
Using new proper motion data from recently published catalogs, we revisit the membership of previously identified members of the Upper Scorpius association. We confirmed 750 of them as cluster members based on the convergent point method, compute their kinematic parallaxes and combined them with Gaia parallaxes to investigate the 3D structure and geometry of the association using a robust covariance method. We find a mean distance of $146\pm 3\pm 6$~pc and show that the morphology of the association defined by the brightest (and most massive) stars yields a prolate ellipsoid with dimensions of $74\times38\times32$~pc$^{3}$, while the faintest cluster members define a more elongated structure with dimensions of $98\times24\times18$~pc$^{3}$. We suggest that the different properties of both populations is an imprint of the star formation history in this region. 
\end{abstract}

\begin{keywords}
open clusters and associations: individual: Upper Scorpius -- Galaxy: kinematics and dynamics -- stars: distances  -- proper motions -- methods: statistical.
\end{keywords}


\section{Introduction}

The Upper Scorpius association, located at a distance of about 145~pc, is the youngest \citep[$\sim5-10$~Myr;][]{Preibisch2002,Pecaut2016} and best-studied subgroup of the Scorpius-Centaurus complex. Despite its close proximity to the Ophiuchus star-forming clouds there are no indications of ongoing star formation activity. The molecular clouds in this region have already been dispersed so that cluster members of different masses are still present and can be easily observed. \citet{deZeeuw1999} investigated the high-mass stellar population of the association ($M\geq 2M_{\odot}$) and identified most of the group members with spectral types B, A and F. Later studies focused on the low-mass content of the association, identifying hundreds of late-type stars and brown dwarfs \citep[see e.g.][]{Preibisch1999,Preibisch2002,Luhman2012,Rizzuto2015,Lodieu2013,Cook2017}.

The age estimate of Upper Scorpius is controversial. While previous works reported an age of $\sim 5$ Myr with very little spread \citep{Preibisch1999,Preibisch2002,Slesnick2008}, more recent studies indicate an older median age (10-11 Myr) with a spread as large as $\sim7$~Myr \citep{Pecaut2012}, with a possible dependence on position \citep{Pecaut2016}, effective temperatures \citep{Rizzuto2016}, or the presence of circumstellar disk \citep{Donaldson2017}. Whether this age spread is real or not \citep[see e.g.][]{Fang2017} remains under debate.

In this context, another important aspect is the intrinsic size of the association and the spread of individual distances which causes the stellar luminosities (and ages) to be over- or under-estimated with respect to the mean distance of the association. \citet{deZeeuw1999} derived a mean distance of $145\pm2$~pc based on trigonometric parallaxes from the \textit{Hipparcos} catalog \citep{HIP97} of 120~bright cluster members. The spread of distances for this sample amounts to about 32 pc, which is consistent with the projected diameter of the association of $\sim14^\circ$ that roughly corresponds to 35~pc. Thus, based on these data we may assume that the bright stars' spatial distribution has a spherical shape \citep[see also][]{Preibisch2008}. Recent studies led by \citet{Fang2017} and \citet{Cook2017} estimated a mean distance of 144 and $145\pm2$ pc, and a spread along the line of sight of about $\pm15$ and $\pm13$~pc, respectively. However,  a complete study of the 3D geometry of the Upper Scorpius association and its dependence with stellar masses is still lacking. This situation will improve dramatically in the near future with the trigonometric parallaxes delivered by the Gaia satellite down to $G\simeq20$ mag.  In this paper we use the first data release of the Gaia satellite (Gaia-DR1) together with the more recent ground-based surveys to investigate the 3D structure and membership of the association.

\section{Sample of Upper Scorpius stars}

The first step in our analysis is to build a complete census of known members in the Upper Scorpius association. We have compiled a list of 1322~stars that were identified in previous studies to be likely members of the association based on youth diagnostics and proper motions \citep{deZeeuw1999,Preibisch1999,Preibisch2002,Rizzuto2011,Luhman2012,Rizzuto2015,Pecaut2016}. Of course, this sample of candidate members will undergo revision as soon as more data from ongoing and upcoming surveys (e.g. Gaia, LSST) become available. In the following, we searched the current databases to access the more recent data available for this sample that will be useful to investigate the structure of the association. 

The Tycho-Gaia Astrometric Solution \citep[TGAS,][]{Lindegren2016} that has just been delivered by Gaia-DR1 provides trigonometric parallaxes and proper motions for only 149~stars. On the other hand, the recently published ``Hot Stuff for One Year'' \citep[HSOY,][]{HSOY} and UCAC5 proper motion catalogs provide proper motion measurements for 1011~stars and 729~stars in our sample, respectively. Both catalogs combine the stellar positions from Gaia-DR1 with ground-based astrometry from the PPMXL \citep{PPMXL} and UCAC4 \citep{Zacharias2013} catalogs, and represent the best present-day compromise for  this work between the number of sources in our sample and proper motion precision. In addition, we also complement these two sources of proper motions with data from the SPM4 \citep{SPM4} catalog for 783~stars. We compared the proper motions of the stars in our sample given by the different catalogs and detected some outliers in UCAC5. Thus, we use the TGAS and HSOY catalogs as main sources of proper motions and complement them with proper motions from UCAC5 and SPM4. Doing so, we find proper motion data for 1104~stars of our initial sample. 

We also searched the literature for radial velocity information using the data mining tools available at the CDS/SIMBAD databases \citep{Wenger2000}. Our search for radial velocities is based on \citet{Barbier-Brossat2000,Nordstrom2004,Gontcharov2006,Torres2006,Kharchenko2007,White2007,Holmberg2007,Chen2011,Dahm2012,Song2012,Kordopatis2013,Malo2014,Mann2016}. We found radial velocities for 146~stars. 

Table~\ref{tab1} lists the data collected for the stars in our sample and the results obtained in the forthcoming analysis. It is available only in electronic form at the CDS.

\section{Kinematic parallaxes and membership analysis}\label{section3}

The small number of stars with measured trigonometric parallaxes (i.e., 11\% of the sample) and the scarcity of radial velocity measurements are the main limitations in this work to investigate the structure and kinematics of Upper Scorpius. In this context, individual parallaxes for comoving members of the association sharing the same space motion can be inferred from the moving-cluster method \citep{deBruijne1999a,Galli2012}. The so-derived kinematic parallaxes are not as precise and accurate as the trigonometric parallaxes from the TGAS catalog, but they still provide useful information to complement the forthcoming analysis. 

To begin with, we take the sample of 1104~stars with known proper motions and perform a 3$\sigma$ clipping on the distribution of both proper motion components to remove obvious outliers. This step reduces the sample to 891~stars. Then, we apply the convergent point search method (CPSM) as described in Sect.~2 of \citet{Galli2017}. The method takes a distance estimate and velocity dispersion of the cluster as input parameters. We use the distance of 145~pc and velocity dispersion of $\sigma_{v}\simeq 1.5$~km/s derived by \citet{deBruijne1999b}. The method is rather insensitive to small variations of these parameters. Doing so, the CPSM selects 750~stars that we consider to be confirmed members of the Upper Scorpius association (see Table~\ref{tab1}). The corresponding convergent point solution is located at $(\alpha_{cp},\delta_{cp})=(95.2^{\circ},-42.8^{\circ})\pm(2.2^{\circ},2.5^{\circ})$ with $\chi_{red}^{2}=0.94$ and correlation coefficient of $\rho=-0.99$.

We note that 103~stars among the confirmed moving group members have trigonometric parallaxes, but only 51~stars exhibit complete data (proper motion, radial velocity and trigonometric parallax). We convert their trigonometric parallaxes into distances using an exponentially decreasing space density prior with length scale $L=1.35$~kpc as described by \citet{Astraatmadja2016}, and use them to calculate the three-dimensional Galactic spatial velocities $UVW$ from the procedure described by \citet{Johnson1987}. Then, we perform an iterative $3\sigma$ clipping in the distribution of the $UVW$ spatial velocities and we end up with 40~stars that define our control sample (see Table~\ref{tab1}). We argue that these stars are secure members of the association based on (i) the membership analysis performed with the  CPSM and (ii) their common space motion as derived directly from proper motions, radial velocities and trigonometric parallaxes. We provide in Table~\ref{tab_vel} the spatial velocity of the Upper Scorpius association as derived from our control sample. 
 
\begin{table}
\renewcommand\thetable{2} 
\centering
\caption{Spatial velocity of the Upper Scorpius association derived from our control sample of 40~stars with complete data. We provide for each velocity component the mean, standard error of the mean (SEM), median, mode and standard deviation (SD). 
\label{tab_vel}}
\begin{tabular}{lcccc}
\hline
&Mean $\pm$ SEM&Median&Mode& SD\\
&(km/s)&(km/s)&(km/s)&(km/s)\\
\hline
$U$&$-5.0\pm 0.1$&-4.9&-4.7&2.2\\
$V$&$-16.6\pm 0.1$&-16.9&-16.7&1.5\\
$W$&$-6.8\pm 0.1$&-6.6&-6.6&1.5\\
\hline
$V_{space}$&$18.8\pm 0.1$&18.8&18.8&1.6\\
\hline
\end{tabular}
\end{table}

We calculate kinematic parallaxes for the 750~stars identified as moving group members in our analysis using the formalism described in Sect.~2.1 of \citet{Galli2017}. We decided to compute the kinematic parallaxes using Eq.~2 of \citet{Galli2017} that is written in terms of the spatial velocity of the cluster. Although this procedure applies to all stars in the sample (including binaries and cluster members without radial velocity measurements), the so-derived parallaxes are less accurate than the results obtained directly from Eq.~1 of \citet{Galli2017} based on the radial velocity of individual stars. To overcome this issue we performed Monte Carlo simulations (with 1000 iterations) by resampling the input parameters (proper motion, spatial velocity of the cluster and angular distance to the convergent point position) in Eq.~2 of \citet{Galli2017} from Gaussian distributions where mean and variance correspond to the observed parameters and their uncertainties. We report for each star in Table~\ref{tab1} the mean and standard deviation of the distribution of simulated kinematic parallaxes. The comparison with the trigonometric parallaxes in our control sample for the sample in common (40~stars) yields a mean difference of -0.02~mas (in the sense ``trigonometric'' minus ``kinematic'') and r.m.s. of 0.5~mas.  The mean error of the kinematic parallaxes derived in this work is 0.8~mas while the mean error of the TGAS trigonometric parallaxes in our sample is 0.5~mas. The systematic errors of 0.3~mas in the TGAS trigonometric parallaxes reported by \citet{Lindegren2016} were added quadratically to the parallax uncertainties in our analysis. We conclude that both samples (trigonometric and kinematic parallaxes) are consistent between themselves within their errors.  

We obtain a mean parallax of $\pi=7.06\pm 0.03$~mas based on the kinematic parallaxes derived in this paper for the 750 stars. This corresponds to a distance of $d=142\pm 1$~pc given a confidence interval of 95\%.  In comparison, the mean parallaxes obtained from the control sample (40~stars) and all TGAS trigonometric parallaxes (103~stars) are, respectively, $\pi=7.09\pm0.09\pm0.30$~mas and $\pi=6.86\pm0.07\pm0.30$~mas. They yield a distance estimate of $d=141\pm4\pm 6$~pc and $d=146\pm3\pm 6$~pc. The second term in our uncertainties refers to the systematic errors of the trigonometric parallaxes in the TGAS catalog \citep[see e.g.][]{Lindegren2016}. We consider the latter distance estimate as our final result, because it is based on a more significant number of cluster members (with measured trigonometric parallaxes) and it will be confirmed in our forthcoming analysis (see Sect.~\ref{section5}).

\section{3D structure}

To investigate the structure of the association we use the stellar parallaxes derived above to calculate the three-dimensional position $XYZ$ of the selected group members in our sample. This reference system has its origin at the Sun, where $X$ points to the Galactic center, $Y$ points in the direction of Galactic rotation and $Z$ points to the Galactic north pole to form a right-handed system. 

To derive the main 3D global  geometrical properties of the Upper Scorpius association, we use a multivariate statistical method based on estimators of the covariance matrix of the 3D spatial distribution. The idea is that the envelope of any 3D distribution of points may be approximated by a Gaussian ellipsoid distribution that is free to rotate around any axes. The center of the ellipsoid is the centroid of the stars of the cluster, and the 3x3 covariance matrix of this distribution of points contains the structural information of its orientation and its major directions of scatter in the 3 dimensions.
Indeed, the eigenvectors of the covariance matrix $C$  constitute an orthonormal system of maximal variance, i.e. it gives the 3 main axes along which the data vary the most. The eigenvalues $\lambda_i$ of the covariance matrix are related to the dispersion of the data along those principal axes. The semi-axes $a_i$ of the ellipsoid are then defined as:  
\begin{equation}
a_i=(\chi^2(\alpha,d)\, \lambda_i)^{1/2}\,, 
\end{equation}
where $\chi^2$ is the chi-squared  function, $d$ is the dimension of the space in which the data are embedded (i.e., $d=3$), and $\alpha$ the quantile that defines the proportion of data that are contained within the ellipsoid.

To derive the vector and the angle of the rotation operation that brings the $XYZ$ coordinate system in the orthonormal system of the optimal ellipsoid, we use the following equation:
\begin{equation} 
C \cdot V = V \cdot \Gamma \, ,
\end{equation}
where $V$ is the matrix whose columns are composed by the 3 eigenvectors  ($\vec{u}_{x^\prime}=\{V_{i1}\}, \, \vec{u}_{y^\prime}=\{V_{i2}\}, \, \vec{u}_{z^\prime}=\{V_{i3}\}; \, i=1,2,3$)  
of the covariance matrix $C$, and $\Gamma$ is a diagonal matrix composed of the corresponding ordered eigenvalues. The matrix $V$ represents the rotation transformation matrix and $\Gamma^{1/2}$ the scale matrix such that we have
\begin{equation} 
C= V\cdot \Gamma^{1/2} \cdot \Gamma^{1/2} \cdot V^{-1}\, .
\end{equation}

In other words we define the linear transformation $T=V \cdot \Gamma^{1/2}$, such that the covariance matrix is defined as $C=T\cdot \, ^{t}T$, where $^{t}T$ is the transpose matrix of $T$. Applying the covariance matrix on any set of white noise will transform the set of white data into the rotated and scaled data that fit the ellipsoid.

The inclination angle $i$ of the most elongated principal axis of the ellipsoid (oriented along the vector $\vec{u}_{X^{\prime}}=\{V_{i1}\}, \,i=1,2,3$) with respect to the Galactic plane is estimated from:
\begin{equation}
i ={\rm atan} \left( V_{31} / (V_{11}^2+V_{21}^2)^{1/2} \right ) \, .
\label{Eq:IncPrincAxe}
\end{equation}

In the standard covariance method, all data are taken into account to compute the covariance matrix $C$, which is fine for a homogeneous distribution. But when the distribution of points is more complex, as for example a compact region with sparse and diffuse tails or the presence of noise and outliers (as in our data), the need for a robust estimator of the covariance matrix is required.  In order to define the main optimal properties of the ellipsoid (location, orientation and scatter), we choose the minimum covariance determinant (MCD) estimator as it is the most robust in presence of a high proportion of outliers \citep{RousseeuwDriessen1999,RousseeuwHubert2011}, when compared for example to the {\textbf robust} Minimum Volume Ellipsoid (MVE) estimator \citep{AelstRousseeuw2009}.  The MCD estimator searches for the ellipsoid with the smallest determinant that covers a specified ``good" fraction of the data. No less than half of the data has to be used. Since the data are not homogeneously distributed, the method may lead to local attractors if the fraction of data is inadequate. We found that taking 80\% as the minimum required fraction of data used in the MCD estimator allows to get rid of the outliers and ensures the uniqueness and stability of the solution.


\begin{table}
\renewcommand\thetable{3}
\centering
\caption{Description of the star sample, the parallaxes and the bootstrap technique used for each numerical experiment.}
\label{Tab:experiment}
\begin{tabular}{ccccc}
  \hline
Experiments   & Sample & N stars  & Parallax & Bootstrap   \\ 
  \hline
$E_{1}$ & $S_1$ & 103   & TGAS         & $B_1$ \\ 
$E_{2}$ & $S_1$ & 103   & Kinematic  & $B_1$ \\ 
$E_{3}$ & $S_2$ & 750   & Kinematic  & $B_1$ \\ 
$E_{4}$ & $S_2$ & 750   & Kinematic  & $B_2$ \\ 
$E_{5}$ & $S_3$ & 647   & Kinematic  & $B_1$ \\ 
$E_{6}$ & $S_3$ & 647   & Kinematic  & $B_2$ \\ 
   \hline
\end{tabular}
\end{table}


We then apply this method to 3 different subsamples of our list of 750~stars considered as cluster members from the CPSM (see previous section): the TGAS stars only, corresponding to the brightest stars ($S_1$, $n_1= 103$ stars); all the stars of our catalog having a kinematic parallax estimate ($S_2$, $n_2= 750$ stars); and finally the stars that have a kinematic parallax estimate but are not in TGAS ($S_3$, $n_3= 647$ stars). From those samples, we perform 6 calculations (see Table~\ref{Tab:experiment}) depending on whether we use the trigonometric or kinematic parallaxes to compute the 3D Galactic coordinates of the stars in sample $S_1$ (respectively experiments $E_1$ and $E_2$), and depending on the bootstrap technique we used for the samples $S_2$ and $S_3$ (experiments $E_3$ to $E_6$).

Indeed, in order to derive the geometrical properties in a statistical and robust way, we implement two different bootstrap iterative  methods (5000 iterations).
The first bootstrap technique ($B_1$) is based on a Monte Carlo sampling applied to the coordinates (right ascension, declination and parallax), taking into account their normally distributed uncertainties to compute their Galactic coordinates. This bootstrap technique aims at evaluating the effect of the individual uncertainties of the star positions on the global structure of the association.
The second bootstrap method ($B_2$) takes into account the fact that the sample sizes are different. When comparing the distribution of the 103 brightest (and therefore most massive assuming they are located approximately at the same distance) stars from sample $S_1$ to the 647 less bright (less massive) stars (sample $S_3$), we need to estimate the possible bias introduced by the different sample size. We thus realize a Monte Carlo sampling of 103 stars within the $S_2$ and $S_3$ samples, to which we apply the $B_1$ bootstrap method to compute the Galactic coordinates. For each of the experiments, we derive the mean center, the lengths and direction of the main axes of the ellipsoid, the inclination of the most elongated axis of the ellipsoid with respect to the Galactic plane  
(see Table~\ref{Tab:USGeomProp_sing}). The 3D spatial study of the Upper Scorpius association has been performed in the {\it R environment} \citep{Rmanual2017} using the {\it MASS} package \citep{MASS2002} for the MCD spatial analysis and the {\it  rgl} \citep{rgl2017} package for the 3D visualization.


\begin{table*}
\renewcommand\thetable{4}
\centering
\caption{Geometrical properties of the Upper Scorpius association obtained in each experiment $E_i$ (see Table~\ref{Tab:experiment}). We provide the distance $D_0$ of the ellipsoid centroid to the Sun, their Galactic cartesian coordinates ($x_{0}$, $y_{0}$, $z_{0}$), the eigenvectors of the covariance matrix (i.e. the vectors  $\vec{u}$ of the principal axes of the ellipsoid, ${X^\prime}$,${Y^\prime}$,${Z^\prime}$), the semi-axes ($a_1,a_2,a_3$) of the ellipsoids, inclination $i$ of the first main principal axis with respect to the Galactic plane and the corresponding uncertainties.}
\label{Tab:USGeomProp_sing}
\begin{tabular}{rccccccccc}
  \hline
 Exp. & $D_0$ & $x_{0}$, $y_{0}$, $z_{0}$ & $\vec{u}_{x^\prime}$ & $\vec{u}_{y^\prime}$ & $\vec{u}_{z^\prime}$ & $a_1$  & $a_2$  & $a_3$  & $i$ \\ 
&(pc)&(pc)&&&&(pc)&(pc)&(pc)&($^{\circ}$)\\ 
  \hline
$E_1$ & 145$\pm$1 & 134.4;-22.2;49.8 & 0.897;-0.140;0.418 & -0.391;0.185;0.901 & -0.203;-0.973;0.111 & 37$\pm$3 & 19$\pm$1 & 16$\pm$1 & 25$\pm$3 \\ 
$E_2$ & 144$\pm$2 & 133.1;-21.7;50.5 & 0.935;-0.022;0.353 & -0.351;0.066;0.934 & -0.044;-0.998;0.054 & 38$\pm$3 & 18$\pm$1 & 14$\pm$1 & 20$\pm$6 \\ 
$E_3$ & 142$\pm$3 & 130.4;-16.3;54.9 & 0.904;-0.097;0.417 & 0.063;0.993;0.095 & -0.424;-0.060;0.904 & 51$\pm$4 & 10$\pm$2 & 6$\pm$1 & 25$\pm$1 \\ 
$E_4$ & 142$\pm$3 & 131.3;-18.7;51.6 & 0.912;-0.112;0.395 & -0.271;0.561;0.783 & -0.309;-0.820;0.481 & 48$\pm$5 & 13$\pm$1 & 10$\pm$1 & 23$\pm$2 \\ 
$E_5$ & 142$\pm$2 & 129.8;-16.5;54.5 & 0.903;-0.096;0.418 & 0.006;0.977;0.213 & -0.429;-0.191;0.883 & 52$\pm$4 & 10$\pm$1 & 6$\pm$1 & 25$\pm$1 \\ 
$E_6$ & 142$\pm$3 & 130.7;-18.3;51.6 & 0.909;-0.113;0.401 & -0.264;0.587;0.765 & -0.322;-0.801;0.504 & 49$\pm$5 & 12$\pm$1 & 9$\pm$1 & 24$\pm$2 \\ 
   \hline
\end{tabular}
\end{table*}

\begin{figure}
\begin{center}
\includegraphics[width=0.50\textwidth]{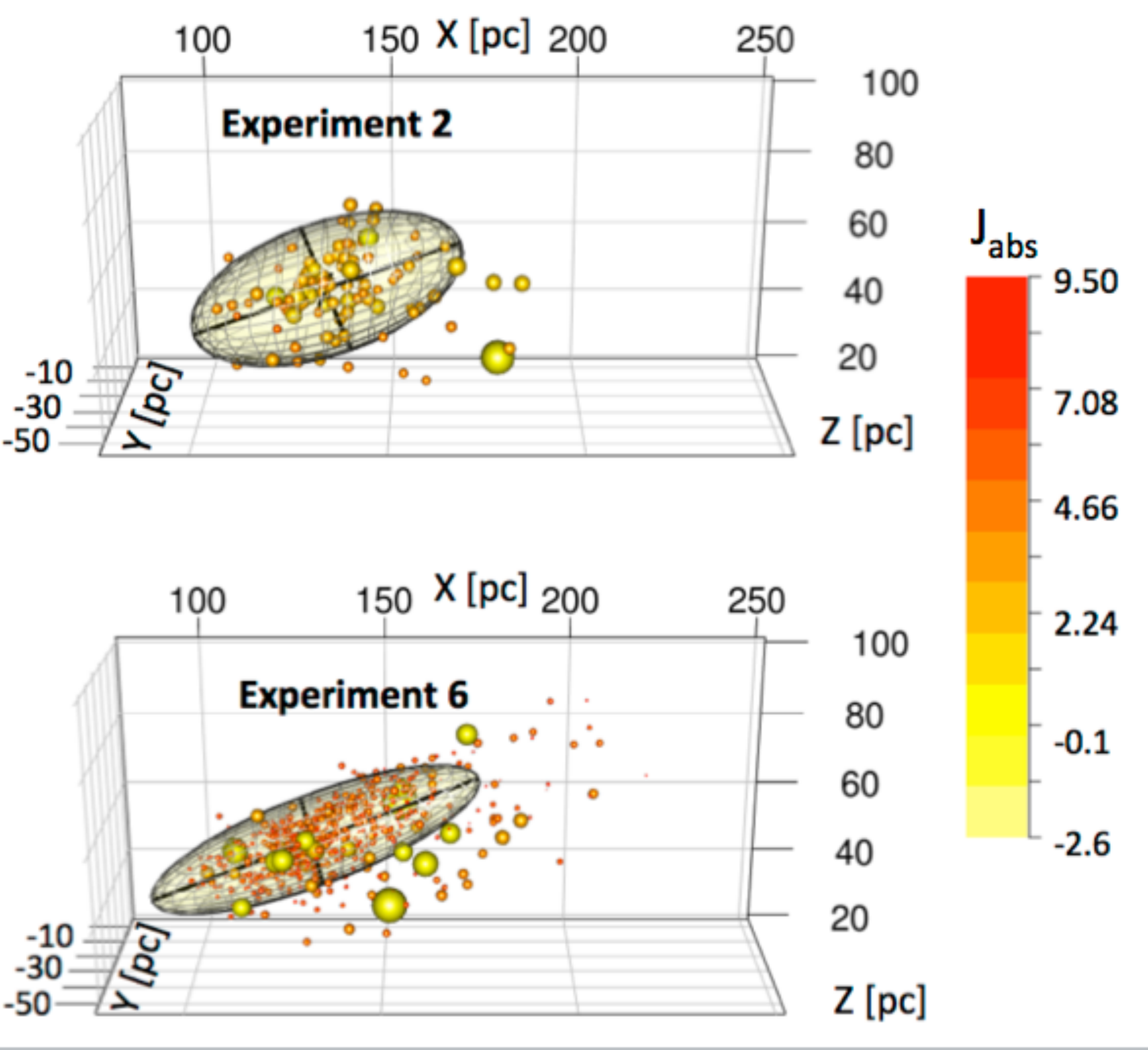}
\caption{
\label{fig:3D_UpSco_1}
3D structure of the Upper Scorpius association obtained for the experiments $E_2$ (upper panel) and $E_6$ (lower panel). The color and size of the points are related to the absolute magnitude of the stars computed from the 2MASS $J$ filter.  
}
\end{center}
\end{figure}

\section{Discussion and conclusion}\label{section5}

Despite the different data sets used in $E_{1}$ and $E_{2}$ (trigonometric and kinematic parallaxes, see Table~\ref{Tab:experiment}) we confirmed that the global 3D distributions are similar. In both cases, the geometry of the $S_1$ sample is a prolate structure elongated mainly along the $X$ axis (oriented towards the Galactic center) with the $E_{1}$ dimensions $\sim 74 \times 38 \times 32$ pc$^3$, and an inclination of $i=20^\circ$-$25^\circ$ with the Galactic plane. The computed centroids of the ellipsoids are about 1 pc away from each other and are located at a distance of $145$ and $144$ pc for $E_1$ and $E_2$ respectively. The average value is consistent with our distance estimate of $146\pm3\pm6$ pc derived in Sect.~\ref{section3} for this sample. We note that intriguingly the inclination $i$ obtained in this work is very close to the one derived for the overall structure of the Gould's belt \citep{PerrotGrenier2003, PalousEhlerova2017}.

We check that no bias is introduced by sample size effect in our analysis
by using the two bootstrap techniques described earlier. $E_3$ and $E_4$ (resp. $E_5$ and $E_6$) experiments give the same results within the standard deviations. We compare the shape of the ellipsoids that we obtain for the stars present in the TGAS catalog $S_1$ sample, corresponding in majority to the brightest ($G<12$ mag) and therefore most massive stars, to the fainter and less massive $S_3$ sample. The ellipsoid associated with the faintest stars ($E_6$ experiment) has a 3D dimension of $\sim 98 \times 24 \times 18$ pc$^3$. The ratio of semi-axes between the ellipsoids $E_{2}$ and $E_{6}$ illustrated in Fig.~\ref{fig:3D_UpSco_1} is 1.3, 0.7 and 0.6. These results may suggest that the formation of the more massive stars took place first, before the formation of the less massive stars, which would allow a slight relaxation for the most massive stars set. This would be seen as a less $E_2$ anisotropic structure compared to the very elongated $E_6$ ellipsoid and favour a real age spread.

The anisotropy of the shape of both ellipsoids suggest that the Upper Scorpius association is not dynamically relaxed. Indeed, using a one-dimensional velocity dispersion of 2.2~km/s for the brightest stars sample (see Table~\ref{tab_vel}) and 11~Myr \citep{Pecaut2012} as the estimated age of the region, the crossing distance is of the order of $\sim25$~pc, which is less than the largest dimension of the ellipsoid. We thus conclude that the shape of the Upper Scorpius association is an imprint of its star formation history. This study represents an important step towards understanding the complex history of the star formation process in this region. The more accurate parallaxes from Gaia-DR2 for the faintest cluster members will allow us to build on this scenario.  

\section*{Acknowledgements}

We thank the referee for constructive comments that helped
us to improve the manuscript. This project has been partly supported by the ``StarFormMapper'' project funded by the European Union's Horizon 2020 Research and Innovation Action (RIA) programme under grant agreement number 687528. P.A.B.G. acknowledges financial support from FAPESP (grants: 2013/04934-8 and 2015/14696-2). This research has made use of the computing facilities of the Laboratory of Astroinformatics at IAG/USP (S\~ao Paulo, Brazil). This research made use of the SIMBAD database operated at the CDS (Strasbourg, France).


\bibliographystyle{mnras}
\bibliography{references} 

\bsp	
\label{lastpage}



\begin{landscape}
\renewcommand\thetable{1} 
\begin{table}
\caption{Properties of the 1322~stars in the Upper Scorpius region. We provide for each star the 2MASS identifier, position, proper motion, radial velocity, kinematic and trigonometric parallax, spatial velocity and membership status based on the CPSM (`Y'$=$yes, `N'$=$no). The CTRL column (`Y'$=$yes, `N'$=$no) indicates the stars included in our control sample. The full table is available online. }\label{tab1}
\tiny{
 \begin{tabular}{lcccccccccccccc}
\hline
2MASS Identifier & $\alpha$& $\delta$  & $\mu_{\alpha}\cos\delta$ & $\mu_{\delta}$ & Ref. &$V_{r}$&Ref.&$\pi_{kin}$&$\pi_{trig}$&$U$&$V$&$W$&CPSM&CTRL\\
& (h:m:s) &($^{\circ}$ $^\prime$ $^\prime$$^\prime$)  & (mas/yr) & (mas/yr)&&(km/s)&&(mas)&(mas)&(km/s)&(km/s)&(km/s)&&\\
\hline

J15302162-2036481	&	15 30 21.63	&	-20 36 48.12	&$	-19.630	\pm	0.176	$&$	-23.430	\pm	0.071	$&	TGAS	&$	-31.3	\pm	0.3	$&	1	&$	7.97	\pm	0.71	$	&$	6.84	\pm	0.39	$&	$	-30.2	_{	-0.9	}^{+	0.9	}$&	$	-14.3	_{	-1.6	}^{+	1.2	}$&	$	-17.8	_{	-1.4	}^{+	1.3	}$&	Y	&	N	\\
J15311722-3349115 	&	15 31 17.23	&	-33 49 11.56	&$	-17.764	\pm	0.985	$&$	-25.071	\pm	0.942	$&	HSOY	&				&		&					&				&							&							&							&	N	&	N	\\
J15315022-3252519 	&	15 31 50.22	&	-32 52 51.92	&$	-17.269	\pm	0.035	$&$	-22.233	\pm	0.020	$&	TGAS	&				&		&					&$	5.95	\pm	0.78	$&							&							&							&	N	&	N	\\
J15321033-2158004	&	15 32 10.33	&	-21 58 00.48	&$	-16.07	\pm	1.99	$&$	-23.58	\pm	1.88	$&	SPM4	&				&		&$	7.46	\pm	0.82	$	&				&							&							&							&	Y	&	N	\\
J15322013-3108337 	&	15 32 20.14	&	-31 08 33.75	&$	-18.845	\pm	0.057	$&$	-22.491	\pm	0.042	$&	TGAS	&$	-1.4	\pm	0.3	$&	1	&					&$	7.75	\pm	0.52	$&	$	-6.3	_{	-0.8	}^{+	0.6	}$&	$	-16.6	_{	-1.6	}^{+	1.1	}$&	$	-4.0	_{	-1.4	}^{+	1.3	}$&	N	&	N	\\
J15325521-1651101 	&	15 32 55.21	&	-16 51 10.18	&				&				&		&$	1.8	\pm	1.1	$&	2	&					&				&							&							&							&	N	&	N	\\
J15330951-2429253 	&	15 33 09.51	&	-24 29 25.39	&$	-24.748	\pm	0.111	$&$	-33.503	\pm	0.055	$&	TGAS	&$	-3.6	\pm	0.6	$&	3	&					&$	11.79	\pm	1.00	$&	$	-6.0	_{	-1.2	}^{+	1.1	}$&	$	-15.4	_{	-2.1	}^{+	1.3	}$&	$	-5.1	_{	-1.8	}^{+	1.6	}$&	N	&	N	\\
J15350863-2532397	&	15 35 08.64	&	-25 32 39.73	&				&				&		&				&		&					&				&							&							&							&	N	&	N	\\
J15351610-2544030	&	15 35 16.10	&	-25 44 03.07	&$	-18.428	\pm	0.042	$&$	-23.482	\pm	0.023	$&	TGAS	&$	-2.54	\pm	0.60	$&	4	&$	7.62	\pm	0.63	$	&$	7.44	\pm	0.62	$&	$	-6.0	_{	-1.3	}^{+	1.1	}$&	$	-17.9	_{	-2.2	}^{+	1.4	}$&	$	-4.7	_{	-1.9	}^{+	1.7	}$&	Y	&	Y	\\
J15354856-2958551	&	15 35 48.56	&	-29 58 55.18	&$	-41.096	\pm	2.370	$&$	-40.788	\pm	2.369	$&	HSOY	&				&		&					&				&							&							&							&	N	&	N	\\

\hline
\end{tabular}
}
\textbf{Radial velocity references.} (1)~\citet{Chen2011}, (2)~\citet{Kharchenko2007}, (3)~\citet{Gontcharov2006}, (4)~\citet{Dahm2012}, (5)~\citet{Torres2006}, (6)~\citet{White2007}, (7)~\citet{Mann2016} and (8)~\citet{Kordopatis2013}. 
\end{table}
\end{landscape}

\end{document}